\documentclass{article}
\usepackage{spconf,amsmath,graphicx}
\usepackage{cite}
\usepackage{epstopdf}
\usepackage{amssymb}
\usepackage{caption}
\usepackage{tabularx}
\usepackage{booktabs}
\usepackage{algorithm}
\usepackage{algorithmic}
\usepackage{multirow}
\usepackage{stfloats}
\usepackage{setspace}
\usepackage{booktabs}
\usepackage{graphicx}
\usepackage{subcaption}

\title{Hierarchical Modeling of Spatial Cues via Spherical Harmonics for Multi-Channel Speech Enhancement}

\name{Jiahui Pan, Shulin He, Hui Zhang, Xueliang Zhang}

\address{
College of Computer Science, Inner Mongolia University, China\\
\texttt{panjiahui@mail.imu.edu.cn, \{cszh,cszxl\}@imu.edu.cn}
}

\begin{document}
\setstretch{0.84}
\maketitle
\begin{abstract}
Multi-channel speech enhancement utilizes spatial information from multiple microphones to extract the target speech. However, most existing methods do not explicitly model spatial cues, instead relying on implicit learning from multi-channel spectra. To better leverage spatial information, we propose explicitly incorporating spatial modeling by applying spherical harmonic transforms (SHT) to the multi-channel input. In detail, a hierarchical framework is introduced whereby lower order harmonics capturing broader spatial patterns are estimated first, then combined with higher orders to recursively predict finer spatial details. Experiments on TIMIT demonstrate the proposed method can effectively recover target spatial patterns and achieve improved performance over baseline models, using fewer parameters and computations. Explicitly modeling spatial information hierarchically enables more effective multi-channel speech enhancement.
\end{abstract}
\section{Introduction}
\label{sec:intro}
Multi-channel speech enhancement is the process of enhancing a target's speech corrupted by background interference using multiple microphones. It is crucial to many applications including, but not limited to, human-machine interfaces \cite{bai2013acoustic}, mobile communication \cite{tan2019real}, and hearing aids \cite{nossier2019enhanced}.
While the problem has been studied extensively, effectively integrating and processing spatial cues remains an open challenge.

Traditional approaches include spatial filtering methods \cite{hoshuyama1999robust,souden2009optimal} that often utilize spatial information from the sound scene, such as the angular position of the target speech and microphone array configuration. These approaches are commonly termed beamforming. Classic beamforming algorithms such as the delay-and-sum beamformer \cite{klemm2008improved}, minimum variance distortionless response (MVDR) \cite{souden2009optimal} beamformer, and super-directivity beamformer \cite{bitzer2001superdirective} can perform well, but their performance depends on reliable estimation of spatial information, which can be challenging to accurately estimate in noisy conditions.

Recently, with the tremendous success of deep learning, speech enhancement has been formulated as a supervised learning problem \cite{wang2018supervised,weninger2015speech}.
In order to effectively utilize spatial information, a common strategy is to combine DNNs with traditional beamforming techniques. Examples include FasNet \cite{luo2019fasnet}, EabNet \cite{li2022embedding}, and MIMO-Unet \cite{ren2021causal}, which have achieved state-of-the-art performance by exploiting the complementary strengths of data-driven deep learning and array signal processing. Specifically, time-frequency (TF) masks can first be predicted by DNNs, then used to determine beamforming weights based on statistical theories such as MVDR \cite{xiao2016study} and generalized eigenvalue (GEV) \cite{heymann2015blstm}. 
However, as the second stage relies purely on statistical theory without considering the mask estimation, pre-estimation errors may severely degrade subsequent beamforming performance \cite{ren2021causal}.
Spatial information is vital for multichannel scenarios. However, without explicit spatial features as input, the above approaches only use multi-channel spectra, relying on networks to learn spatial information implicitly. Other strategies adopt the reference channel magnitude and inter-channel phase differences (IPD) as network inputs \cite{gu2019end}. However, using only phase information to convey channel correlation leads to significant information loss.

Fortunately, the spherical harmonic domain representation offers an innovative approach to decompose the soundfield into spatial basis functions defined on a sphere, concisely depicting its spatial characteristics \cite{kumar2016near,varanasi2020deep}. 
Recent advancements have incorporated spherical harmonic domain processing in various applications including DOA estimation \cite{weng2023doa}, \cite{ben2021binaural} and localization \cite{nadiri2014localization}.
The spherical harmonic coefficients(SHCs) can be deduced from microphone signals via the SHT. This separation of spatial and spectral components provides benefits including \cite{rafaely2015fundamentals}: enabling detailed spatial analysis, circumventing issues with covariance matrices, and allowing soundfield rotation. 
If fully utilizing the spatial cues encoded in SHCs, this may compensate for the limited spatial modeling in mainstream multichannel speech enhancement approaches, thereby improving performance and robustness.

In this paper, we propose hierarchical modeling of spatial cues via spherical harmonics for multi-channel speech enhancement. The approach structurally estimates clean SHCs from noisy SHCs. Lower order harmonics capturing broader patterns are estimated first, then recursively combined with higher orders to predict finer spatial details.
Experiments demonstrate the proposed method can effectively recover target spatial patterns and achieve improved performance over baseline models, while using fewer parameters and computations. 
The results highlight the benefits of leveraging spherical harmonic domain processing and hierarchical prediction to improve multi-channel speech enhancement performance.
Our main contributions are: 
\begin{itemize}
    \item Introducing SHT to enable structured spatial soundfield analysis for multi-channel speech enhancement.
    \item Proposing a divide-and-conquer framework to hierarchically predict coefficient orders, providing gains in both accuracy and efficiency.
\end{itemize}

\begin{figure*}[ht]
	\centering
	\includegraphics[width=1.0\textwidth]{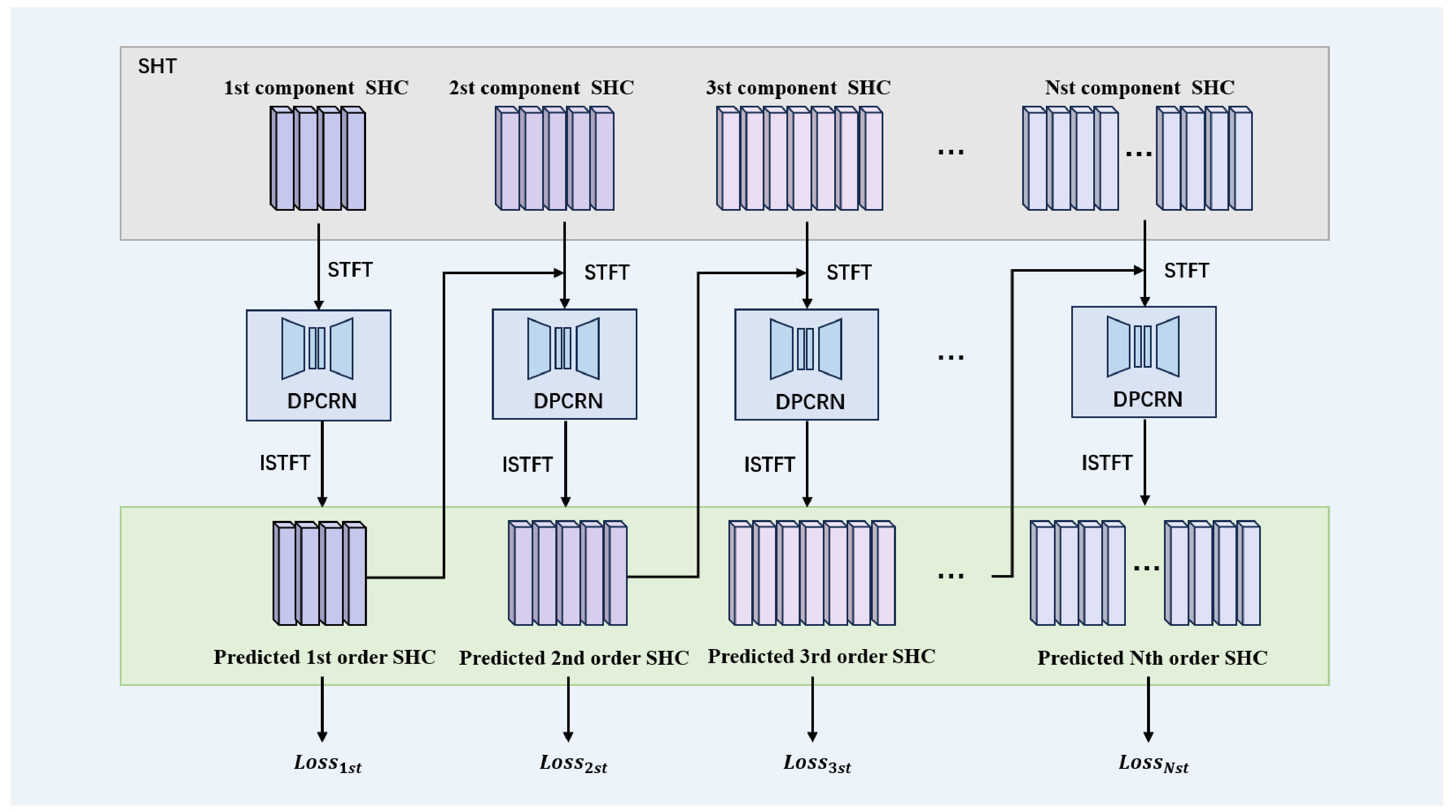}
	\caption{Overview of hierarchical modeling of spatial cues via spherical harmonics.}
	\label{fig:fig1}
\end{figure*}
\section{Method}
The proposed method leverages the hierarchical nature of SHCs to progressively model the spatial information (as illustrated in Fig.\ref{fig:fig1}). Spherical harmonics enable spectral decomposition into coarser components characterized by lower-order coefficients, with finer details encoded in higher orders. The method incrementally predicts SHCs of increasing order from the multi-channel mixed input using a neural network.
Specifically, given the SHCs up to order N, we separate them into orders 1 to N-1 and the Nth order component. 
First, the first-order SHC components are input to a neural network to predict the first-order SHCs of the target speech. The predicted first-order SHCs are then combined with the second-order SHC components to form mixed second-order SHCs. These are input to a second neural network to predict the second-order SHCs. 
This progressive process continues, with each subsequent neural network predicting the SHCs of its order. The inputs to each network are the predicted lower-order SHCs combined with the original SHCs of the same order. In this way, each network leverages both the original and predicted low-order SHC components to estimate the higher-order SHCs in a hierarchical manner. This allows the model to capitalize on the interdependencies between low- and high-order SHCs for improved prediction performance.

\subsection{Spherical Harmonic Representations}
By calculating the coefficients of spherical harmonics, the received speech signal at a specific point on the sphere surface can be estimated.
The SHT enables a Fourier-related spectral analysis on the sphere or in 3D space, expressing signals in terms of SHCs at different orders. In this work, the input multichannel mixed speech is first transformed into the SH domain up to order N, yielding the SHCs. 
Let $\mathbf{r}_i=(r_i\cos\phi_i\sin\theta_i, r_i\sin\phi_i\sin\theta_i, r_i\cos\theta_i)^T$ denote the position of the $i$-th microphone of the array, where $r_i$ represents the distance of the $i$-th microphone to the center of the array. The azimuth $\phi_i$ is measured counterclockwise from the x-axis, and the elevation angle $\theta_i$ is measured downward from the z-axis. The signal received by the $i$-th microphone in the frequency domain can then be expressed as:
\begin{equation}
p_i(k)=\sum_{l=1}^L v_i\left(k, \Psi_l\right) s_l(k)+n_i(k),
\label{p_i(k)}
\end{equation}
where $v_i\left(k, \Psi_l\right)$ denotes the steering vector of the $i$-th microphone associated with the $l$-th plane wave; $ s_l(k) $ is the complex amplitude of the $l$-th plane wave, and $ n_i(k) $ is the noise received by the $i$-th microphone. 

According to the Fourier acoustic principle, the sampled sound pressure ${p}(k, r)$ and its spherical harmonic domain representation ${p}_{n m}(k, r)$ at angle $(\theta, \phi)$ can be expressed as:
\begin{equation}
p(k, r)=\sum_{n=0}^{\infty} \sum_{m=-n}^n p_{n m}(k, r) Y_n^m(\theta, \phi),
\label{pkr}
\end{equation}
\begin{equation}
Y_n^m(\theta, \phi)=\sqrt{\frac{(2 n+1)}{4 \pi} \frac{(n-m) !}{(n+m) !}} P_n^m(\cos \theta) e^{i m \phi},
\end{equation}
where $(.)!$ represents the factorial function. $Y_n^m(\theta, \phi)$ denotes the spherical harmonic functions of order $n$ and degree $m$, and $P_n^m$ is the normalized associated Legendre polynomial.
The spherical harmonic function $P_n^m(\cos \theta)$ captures the dependency on the elevation angle $\theta$, while the complex exponential term $e^{i m \phi}$ captures the dependency on the azimuth angle $\phi$.

As explained in \cite{rafaely2015fundamentals}, the coefficients $p_{nm}$ diminish for $kr$ in a range smaller than N and can therefore be neglected. Hence, Eq.\ref{pkr} can be approximated for an appropriate finite order N:
\begin{equation}
p(k, r) \cong \sum_{n=0}^N \sum_{m=-n}^n p_{n m}(k, r) Y_n^m(\theta, \phi),
\end{equation}
where N is the truncation order, $p(k, r)$ denotes the time-dependent amplitude of the sound pressure in free three-dimensional space,
$Y_n^m = \left[Y_0^0, Y_1^{-1}, Y_1^0, Y_1^1,\cdots, Y_N^N\right]$,  
$p_{nm}(k,r)$ are the weights known as coefficients of the SHT, 
$k = 2\pi f/c$ is the wave number, 
$f$ is the frequency, 
and $c$ is the speed of sound in air. 
The spherical harmonic coefficients $p_{nm}(k,r)$ are defined as \cite{rafaely2015fundamentals}:
\begin{equation}
p_{n m}(k, r)=\int_0^{2 \pi} \int_0^\pi p(k, \mathbf{r})\left[Y_n^m(\theta, \phi)\right]^* \sin (\theta) \mathrm{d} \theta \mathrm{d} \phi,
\end{equation}
where $(.)^*$ denotes complex conjugation. To satisfy the far-field condition, the distance $d$ between the sound source and the center of the microphone array must exceed $8r^{2}f/c$\cite{meyer2001beamforming}, where r is the array radius. This ensures negligible wavefront curvature effects. For $n \leq N$, the spherical harmonic coefficients $p_{nm}(k,r)$ can be obtained as:
\begin{equation}
p_{n m}(k, r) \cong \frac{4 \pi}{I} \sum_{i=1}^I p\left(k, \mathbf{r}_i\right)\left[Y_n^m\left(\theta_i, \phi_i\right)\right]^*,
\end{equation}
where $\mathbf{r}_i=\left(r, \theta_i, \phi_i\right)$ is the location of the i-th physical microphone and I is the number of physical microphones.

\subsection{Progressive Spherical Harmonic Prediction}
The obtained SHCs of the mixed speech are partitioned into incremental groups based on order. The first group comprises the first-order coefficients, corresponding to spherical harmonic basis functions $Y=\left[Y_{0}^{0}, Y_{1}^{-1}, Y_{1}^{0}, Y_{1}^{1}\right]$ in $Y_{n}^{m}$ notation. The second group contains the second-order SHCs excluding the first-order, mapping to basis functions $Y=\left[Y_{2}^{-2}, Y_{2}^{-1}, Y_{2}^{0}, Y_{2}^{1}, Y_{2}^{2}\right]$ in $Y_{n}^{m}$ notation, and so on. This hierarchical partitioning of the spherical harmonic spectrum into distinct order-based components is core to the proposed processing framework.

A progressive neural network architecture is proposed, where each subnet predicts the SHCs of the target speech for a specific order range from the corresponding mixed speech SHCs. The first network takes the first group of SHCs and predicts the 1st order target coefficients. Its output is combined with the second group of SHCs to form the mixed 2nd order coefficients, which are passed to the second network to predict the 2nd order target. This continues progressively until the final network predicts all Nth order SHCs to output a fully enhanced representation.
Each subnet utilizes a DPCRN structure with STFT preprocessing. The loss function is the mean squared error (MSE) between the predicted and true SHCs for each order group. By modeling the residual error at each stage, the network focuses on iteratively refining the spherical harmonic spectral estimate.
\begin{equation}
\text{Loss} = Loss_{1st}+ Loss_{2st}+ Loss_{3st} +\cdots+Loss_{Nst}.
\end{equation}

\section{EXPERIMENTS and Results}
\subsection{Datasets and Evaluation Metrics}
The training data are synthesized by convolving multi-channel room impulse responses (RIRs) \cite{allen1979image} with diverse speech segments from the TIMIT dataset \cite{zue1990speech}. The clean TIMIT segments are split into exclusive training, validation, and testing subsets. Noise segments from the DNS-Challenge corpus are used for training and validation, while testing employs noises from NOISEX-92 \cite{varga1993assessment} and CHiME3 \cite{barker2015third} cafe recordings.
In the data generation phase, relying on a uniform circular array comprising 16 omnidirectional microphones. The array radius is 0.035 m, with a random placement inside the room, while maintaining a source-to-array center distance of 1 m. The generated RIRs pertain to a room with dimensions of $6 \times 5 \times 4 \mathrm{~m}^{3}$, characterized by a variety of SNR and reverberation time RT60 values. SNR values range from -10 dB to 10 dB, while RT60 values span from 0.2 seconds to 1.0 second. Overall, around 24,000 and 2,600 multichannel reverberant noisy mixtures are generated for training and validation, respectively. 
For the purpose of evaluation, we define five distinct SNR levels: -10dB, -5dB, 0dB, 5dB, and 10dB. In addition, we explore nine distinct T60 values, ranging from 0.2s to 1.0s, with intervals of 0.1s. This comprehensive configuration results in the generation of 350 pairs for each specific case.

In this paper, perceptual evaluation of speech quality (PESQ)\cite{rix2001perceptual} and short-time objective intelligibility (STOI)\cite{taal2011algorithm} are chosen as the major objective metrics to evaluate the enhancement performance of different models.
PESQ rates speech quality on a scale from -0.5 to 4.5, while STOI gauges speech intelligibility on a scale of 0 to 100. Improved scores in both metrics reflect better performance.
\begin{table}[t]
\centering
  \caption{Comparisoins if different approaches in Params and FLOPs.}
  \label{tab:hybrid}
    \begin{tabular}{@{}ccc@{}}
        \toprule
        Method & \#Params(M) & FLOPs(G) \\ 
        \midrule
        DPCRN & 5.42 & 21.56 \\
        Proposed & \textbf{3.32} & \textbf{14.06} \\ \bottomrule
    \end{tabular}
\end{table}
\subsection{Experiment Setup}
In our model, we follow the structure of DPCRN \cite{le2021dpcrn}. The channel number of the convolutional layers in the encoder is \{32, 32, 32, 64, 128\}. The kernel size and the stride are respectively set to \{(5,2), (3,2), (3,2), (3,2), (3,2)\} and \{(2,1), (2,1), (1,1), (1,1),(1,1)\} in frequency and time dimension. All the Conv-2D and transposed Conv-2D layers are causally computed. The model includes two DPRNN modules, each containing RNNs with a 128-dimensional hidden state.
The order N of the SHT is set to 4. This generates an $(N+1)^2 \times I$ coefficient matrix, where I is the number of microphones (16 in this work). The dimensions of the corresponding four SHC components are \{(4,16), (5,16), (7,16), (9,16)\}.

The audio is sampled at 16 kHz with a 25 ms window length and 12.5 ms hop size. An FFT length of 400 is used with a sine window applied prior to FFT and overlap-add. The input features are 201-dim complex spectra. Adam optimization is employed with an initial learning rate of 1e-3, halved if validation loss does not decrease for 2 consecutive epochs. Models are trained for 60 epochs.

\begin{table*}[t]
\centering
  \caption{PESQ results comparing proposed models with baselines.}
  \label{tab:table1}
  \resizebox{\linewidth}{!}{
        \begin{tabular}{ccccccccccccccccccc}
           \toprule
           \multirow{2}{*}[-2pt]{\textbf{Methods}} &
           \multicolumn{6}{c}{\textbf{0.2s}} & \multicolumn{6}{c}{\textbf{0.6s}} & \multicolumn{6}{c}{\textbf{1.0s}} \\
           \cmidrule(r){2-7} 
           \cmidrule(l){8-13}
           \cmidrule(l){14-19}
           & -10dB & -5dB & 0dB & 5dB & 10dB & avg. & -10dB & -5dB & 0dB & 5dB & 10dB & avg. & -10dB & -5dB & 0dB & 5dB & 10dB & avg.\\
           \midrule
        Unprocessed
             & 1.25 & 1.34 & 1.48 & 1.76 & 2.13 & 1.59 
             & 1.25 & 1.31 & 1.42 & 1.61 & 1.80 & 1.48 
             & 1.24 & 1.29 & 1.35 & 1.48 & 1.59 & 1.39\\
        GCRN
             & 1.32 & 1.47 & 1.66 & 1.96 & 1.26 & 2.21 
             & 1.31 & 1.44 & 1.58 & 1.80 & 1.97 & 1.99 
             & 1.29 & 1.38 & 1.48 & 1.63 & 1.73 & 1.84\\
        DPCRN
             & 1.41 & 1.69 & 2.04 & 2.50 & 2.88 & 2.10 
             & 1.38 & 1.59 & 1.86 & 2.18 & 2.39 & 1.88 
             & 1.33 & 1.51 & 1.72 & 1.94 & 2.08 & 1.72\\
        \midrule
        \textbf{Proposed}		
             & \textbf{1.51} & \textbf{1.80} & \textbf{2.16} & \textbf{2.62} & \textbf{2.94} & \textbf{2.21} 
             & \textbf{1.46} & \textbf{1.69} & \textbf{1.99} & \textbf{2.32} & \textbf{2.51} & \textbf{1.99} 
             & \textbf{1.39} & \textbf{1.61} & \textbf{1.85} & \textbf{2.10} & \textbf{2.25} & \textbf{1.84} \\
        \bottomrule
        \end{tabular}
    }
\end{table*}

\begin{table*}[t]
\centering
  \caption{STOI results comparing proposed models with baselines.}
  \label{tab:table2}
  \resizebox{\linewidth}{!}{
        \begin{tabular}{ccccccccccccccccccc}
           \toprule
           \multirow{2}{*}[-2pt]{\textbf{Methods}} &
           \multicolumn{6}{c}{\textbf{0.2s}} & \multicolumn{6}{c}{\textbf{0.6s}} & \multicolumn{6}{c}{\textbf{1.0s}} \\
           \cmidrule(r){2-7} 
           \cmidrule(l){8-13}
           \cmidrule(l){14-19}
           & -10dB & -5dB & 0dB & 5dB & 10dB & avg. & -10dB & -5dB & 0dB & 5dB & 10dB & avg. & -10dB & -5dB & 0dB & 5dB & 10dB & avg.\\
           \midrule
        Unprocessed
             & 43.94 & 53.60 & 64.12 & 74.94 & 84.66 & 64.25
             & 40.16 & 48.69 & 57.75 & 67.09 & 75.24 & 57.79  
             & 36.45 & 43.05 & 50.66 & 57.69 & 64.37 & 50.44\\
        GCRN
             & 45.32 & 57.90 & 68.60 & 78.49 & 85.58 & 67.18
             & 41.98 & 53.60 & 63.40 & 72.40 & 78.85 & 62.05
             & 38.08 & 47.95 & 57.52 & 65.44 & 71.51 & 56.10\\
        DPCRN
             & 50.69 & 64.33 & 75.94 & 84.62 & 90.76 & 73.27 
             & 46.92 & 60.44 & 71.10 & 79.65 & 85.48 & 68.72
             & 42.70 & 55.69 & 66.21 & 74.46 & 80.25 & 63.86\\
        \midrule
        \textbf{Proposed}		
             & \textbf{53.96} & \textbf{67.69} & \textbf{78.67} & \textbf{86.50} & \textbf{91.51} & \textbf{75.67} 
             & \textbf{49.33} & \textbf{63.23} & \textbf{73.54} & \textbf{81.40} & \textbf{86.39} & \textbf{70.78} 
             & \textbf{44.76} & \textbf{58.01} & \textbf{68.94} & \textbf{76.59} & \textbf{81.78} & \textbf{66.02} \\
        \bottomrule
        \end{tabular}
    }
\end{table*}

\subsection{Results and Discussions}
\label{ssec:Results}

The proposed system is compared to GCRN \cite{tan2019learning} and DPCRN models. Unlike the above methods without explicit spatial features that only use STFT inputs, the proposed model operates in the spherical harmonic domain to leverage spatial information.
GCRN utilizes convolutional recurrent networks for complex spectral mapping to enhance both magnitude and phase, relying on implicit spatial learning. DPCRN combines CNN local pattern modeling with DPRNN long-term sequence modeling for time-frequency speech enhancement, also based only on STFT inputs.
The proposed model adopts DPCRN as the subnet within a structured spherical harmonic framework to capture spatial information. This hierarchical approach models spatial cues by transforming the multi-channel speech into SHCs and progressively estimating the target SHCs by order.

Table 1 compares the computational complexity and number of parameters for DPCRN and the proposed system. our method requires only 3.32M of the parameters and 14.06G of the FLOPs compared to DPCRN. This highlights the efficiency benefits of the proposed progressive order-wise prediction framework.

As shown in Tables 2 and 3, for a reverberation time (RT60) of 0.2s, the proposed system obtains superior performance across all SNRs, with average PESQ gains of 0.11 and average STOI improvements of 2.4\% compared to DPCRN methods. Notably, the proposed system provides more substantial benefits at lower SNRs.
For RT60=0.6s, the proposed method maintains significant gains over DPCRN, with average PESQ increases of 0.11 and STOI boosted by up to 2.06\%. This indicates that our method of structured modeling is more robust to reverberation compared to the baseline method.
For the most reverberant case with RT60 of 1.0s. The improvement of our method compared to GCRN and DPCRN is still maintained, with the proposed system achieving average PESQ gains of 0.12 and STOI improved by 2.16\%. This validates the efficacy of spherical harmonic modeling for multi-channel speech enhancement under severely reverberant conditions.

In summary, the experimental results validate the benefits of the proposed method over baselines for multi-channel speech enhancement. The hierarchical spatial filtering in the proposed system provides increased accuracy in terms of PESQ and STOI metrics across different reverberation times. Notably, the gains are maintained even for highly reverberant conditions with RT60 of 1.0s. The structured modeling also enables computational efficiency. Together, these results highlight the importance of efficient spherical harmonic modeling and progressive order-wise prediction for robust and accurate multi-channel speech enhancement, especially in acoustically challenging environments.

\section{CONCLUSION}
\label{sec:print}
In this work, we have proposed a progressive prediction approach for multi-channel speech enhancement in the spherical harmonic domain. The key idea is hierarchical estimation of lower order coefficients first, followed by recursive prediction of higher orders conditioned on the lower orders. This allows focusing model complexity on successive spatial frequency bands for more efficient optimization. Experiments were conducted for varied reverberation times and input SNRs. The results demonstrate the proposed technique consistently achieves superior speech enhancement over DPCRN networks operating on microphone signals directly, in terms of PESQ and STOI metrics. This highlights the benefits of structured spherical harmonic modeling and progressive order-wise prediction.

\section{Acknowledgements}
This work was partly supported by the China National Nature Science Foundation (No. 61876214, No. 61866030).
\newpage

\small
\setstretch{0.8}
\bibliographystyle{IEEEbib}
\bibliography{strings,refs}

\end{document}